\documentstyle[12pt,epsfig]{article}
\def\eqn{\begin{equation}}
\def\enn#1{\label{#1} \end{equation}}
\def\eq{\[}
\def\en{\]}
\def\r#1{(\ref{#1})}

\def\f#1{\ref{#1}}

\begin{document} {
\setlength{\baselineskip}{12pt}

\pagestyle{plain}
\title{Boundary processing for Monte Carlo Simulations in the Gap-Tooth Scheme.
}

\author{C. W. Gear\thanks{NEC Research Institute, retired}
\and Ioannis G. Kevrekidis\thanks{Princeton  University, Dept of Chemical
Engineering.}}

\maketitle

\begin{abstract}

This note reports on a scheme for interpolating the boundary
conditions between non-adjacent modeling regions when the model is
based on Monte-Carlo computations of a collection of particles.  The scheme
conserves particles in a natural way, and thereby can be made to
conserve other quantities.

\end{abstract}


\section{Introduction}

In Kevrekidis's gap-tooth scheme\cite{gaptooth} which permits
macro-scale computation over large regions from microscopic models
computed over small regions (the teeth), it is necessary to generate
the boundary conditions for each tooth from the solutions in
neighboring teeth by a suitable-order interpolation.

The gap-tooth scheme can be applied to initial-boundary values
problems for evolutionary PDEs.  In it, space is partially ``tiled'' by non-adjacent
boxes, called {\em teeth}, as shown in Figure \f{f1} for one space
dimension.  The time-dependent, microscopic computational model is solved only
inside the teeth, while the solution between the teeth is obtained by
interpolation between the teeth based on smoothness assumptions.

A solution of the model inside the teeth requires appropriate boundary
conditions on the sides of each tooth.  {\rm Edge} teeth (those
adjacent to a boundary of the actual problem) have part of their
boundary specified from the problem, as shown in Figure \f{f1}.  The
other boundaries are the ones for which the conditions must be generated. 

\begin{figure}[p]
\centerline{\psfig{figure=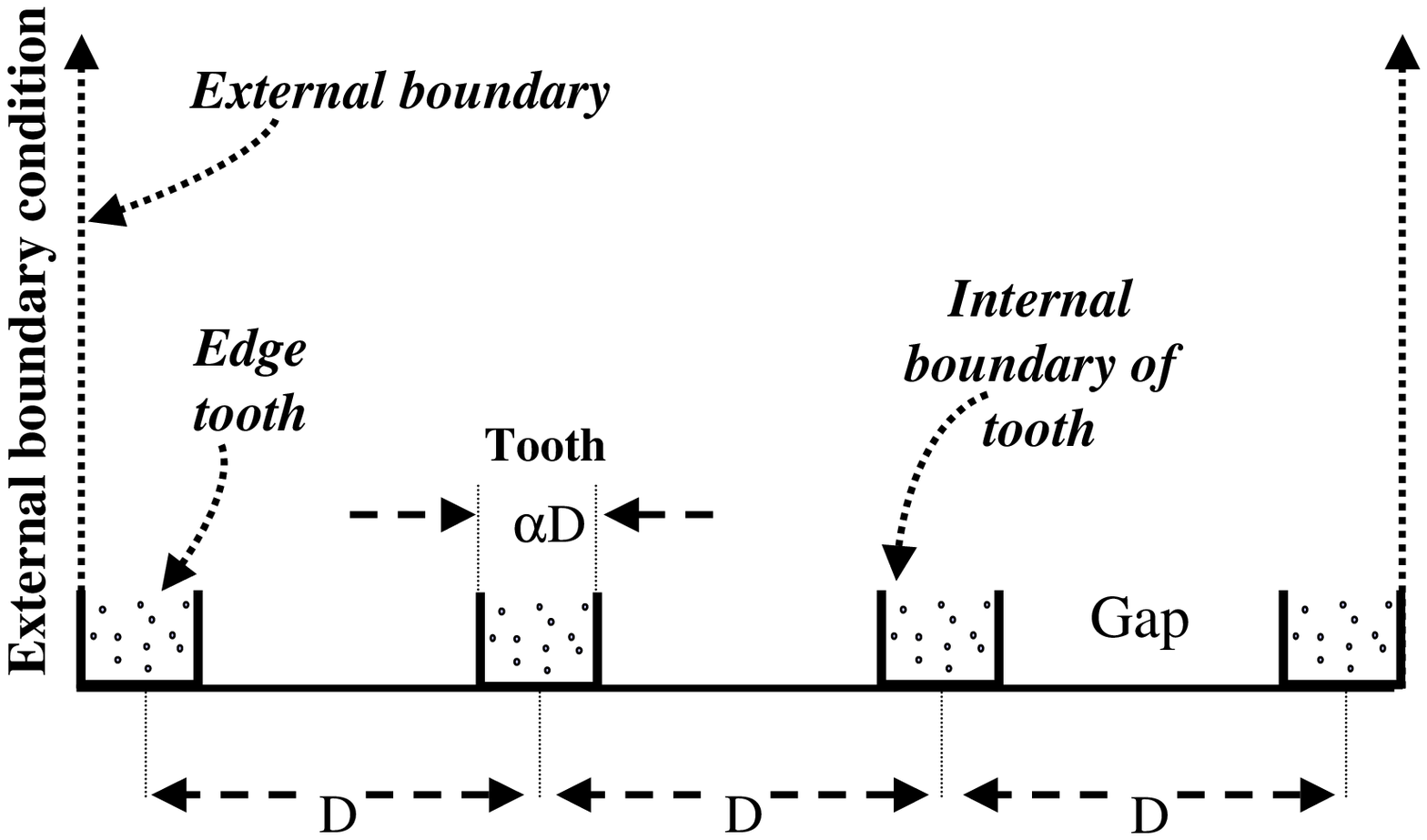,height=3.5in,height=3.5in}}
\vspace{0.1in}
\caption{Teeth partially tiling a region of space}
\end{figure}\label{f1}

Depending on the model used inside the teeth, the variables
interpolated between the teeth might be the same variables, or might
be a {\em restriction} of them to a lower-dimensional set of
variables, the {\em coarse} variables, as described in
\cite{eqfr}.  If the variables inside the teeth are interpolated
between teeth, then the results of those interpolations can be used
directly for the boundary conditions (e.g., the fluxes might be
interpolated, as illustrated in \cite{gaptooth}).  

However, the main point
of modeling over the small teeth is to use microscopic models in
their interior (such as
particle models of fluids or reacting mixtures) in which case the
restriction to the macroscopic (coarse) variables interpolated between the
teeth is to low-order
moments of the solution in each cell, such as density, etc.
If the boundary conditions have to be generated from these
interpolants, it is necessary to use an appropriate computational
interface at the boundary between the restricted coarse variables outside the
teeth and the internal microscopic variables inside the teeth.

In this note we propose a scheme for interpolating boundary conditions
in Monte-Carlo-like particle simulations that is particle-based and avoids the
restriction operation and the resulting boundary interface issues.  The
restriction operation is then necessary to compute the macroscopic
description but not for the far-more-frequently computed boundary conditions..

\section{Linearly-interpolated Particle Boundary Conditions}

When we simulate a system in terms of interacting ``particles,'' each
with a spatial position and possibly additional state variables (such
as velocity, charge, etc.) over a continuum in space, the ``data''
crossing any ``boundary'' between two spatial regions (i.e. any
arbitrary $n-1$ dimensional surface diving the $n$-dimensional space
into two different regions) is a ``discrete flux'' consisting of
particles crossing that boundary from time to time in either
direction, carrying with them their additional state variables such as
velocity and so on.  In macroscopic variable we will typically
characterize these in terms of  the total flux, or velocity, or
pressure, or some combination of these and other moments of particle state variables.
If we are performing a microscopic simulation over the whole of the
spatial domain, there is never any need to ``map'' between the
microscopic boundary conditions and the macroscopic variables except
possibly at the external boundaries where the microscopic boundary
conditions might be handled by such devices as absorbing boundaries,
reflecting boundaries, or periodic boundaries.  However, when we use
the gap-tooth scheme, each additional tooth introduces $2^n$
additional internal boundaries (where $n$ is the space dimension).

In a particle simulation, we can identify two fluxes for each particle
type at a boundary - the outgoing flux of particles leaving the tooth,
and the incoming flux of particles entering the tooth.  The outgoing
flux is naturally generated by the microscopic simulation - in each
simulation time step some number of particles inside a tooth cross a
boundary and are ``lost to that tooth.''  The incoming flux has to be
generated at a boundary as an implementation of the boundary condition
for the microscopic simulation in each tooth.  Figure
\f{f2} shows two neighboring teeth and some of the incoming and
\begin{figure}[p]
\centerline{\psfig{figure=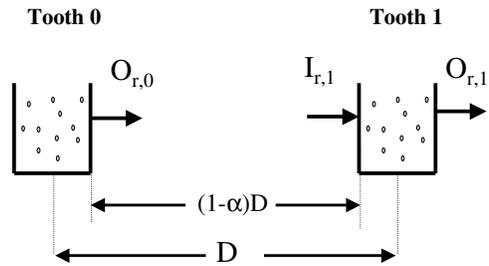,height=3.5in,height=3.5in}}
\vspace{0.1in}
\caption{Input and output right-going fluxes from teeth}
\end{figure}\label{f2}
outgoing fluxes.  We have named the outgoing flux from tooth $i$ as
${\rm O}_{d,i}$ where $d$ is $r$ or $l$ for right-going or left-going
flux, respectively.  (In $n$ space dimensions there would be $2n$
outgoing fluxes from each tooth.)  The incoming fluxes are named ${\rm
I}_{d,i}$.  If there were no gaps (i.e., $\alpha = 1$) the unknown
incoming flux in a particular direction to one tooth would simply be
the known outgoing flux in the same direction from the adjacent tooth,
i.e.,
\eq
{\rm I}_{r,i+1} = {\rm O}_{r,i}
\en
and 
\eq
{\rm I}_{l,i} = {\rm O}_{r,i+1}
\en
When we have a gap, it seems natural to interpolate for ${\rm
I}_{r,i}$ from the nearest ${\rm O}_{r,j}$, that is,
from ${\rm O}_{r,i-1}$ and ${\rm O}_{r,i}$, and similarly for the flux
in the other direction(s).  Using linear interpolation, we have:
\eq
{\rm I}_{r,i} = \alpha {\rm O}_{r,i-1} + (1 - \alpha){\rm O}_{r,i}
\en
and
\eq
{\rm I}_{l,i} = \alpha {\rm O}_{l,i+1} + (1 - \alpha){\rm O}_{l,i}
\en
where $\alpha$ is the ratio of the tooth width to the center spacing
of the teeth.  (In this note, we will discuss only equally-spaced,
equally-sized teeth, although a practical code will almost certainly
need to change the spacing according to the solution behavior.)

In a particle simulation, the outgoing flux is not a real-valued
quantity but, depending on the form of the simulation, either a
discrete-valued quantity - corresponding to the number of particles
that have crossed the boundary in a simulation step if the simulation
proceeds in a sequence of prescribed-length time steps, or a series of
``impulse'' functions - if the simulation is ``event-oriented'' and
proceeds to the next event (which could be a particle-particle
interaction or a boundary crossing).  So what do we mean by an
interpolation between such variables?  Looking at Figure 3 we
note that we have the following expressions for the inputs
\begin{figure}[p]
\centerline{\psfig{figure=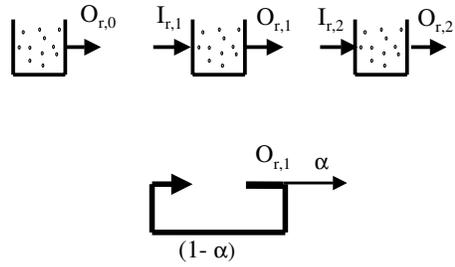,height=3.5in,height=3.5in}}
\vspace{0.1in}
\caption{Linear interpolation of input fluxes by redirection of output fluxes}
\end{figure}\label{f3}
${\rm
I}_{r,1}$ and  ${\rm I}_{r,2}$:
\eq
{\rm I}_{r,1} = \alpha {\rm O}_{r,0} + (1 - \alpha){\rm O}_{r,1}
\en
\eq
{\rm I}_{r,2} = \alpha {\rm O}_{r,1} + (1 - \alpha){\rm O}_{r,2}
\en
Thus, the $\alpha$ of the output ${\rm O}_{r,1}$ contributes to ${\rm
I}_{r,2}$ and $(1 - \alpha)$ of it contributes to ${\rm I}_{r,1}$.  
In other words, we can interpret these equations in a stochastic sense
and {\em send} $\alpha$ of the
particles leaving a tooth to the right to the left end of the right
neighbor, and the remaining $(1 - \alpha)$ back into the left end of
the itself, as shown in the lower part of Figure 3.  The
left-going outputs can be treated similarly.  Thus boundary
conditions for the internal teeth boundaries are handled directly at the
particle level.  We call this a {\em flux redirection scheme}.  This
one is based on linear interpolation.  Note that this approach
preserves the number of 
particles - and will also preserve any other conserved values carried
by the particles.

The way in which the external boundaries are handled necessarily
depends on the specified boundary conditions.  If, for example, an
external boundary is reflecting, one way to handle it is to place an edge
tooth centered on the boundary as shown in Figure 4.  Only the
 \begin{figure}[p]
\centerline{\psfig{figure=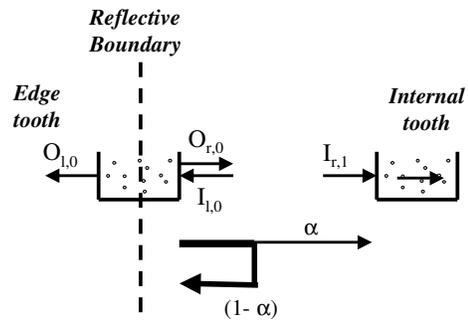,height=3.5in,height=3.5in}}
\vspace{0.1in}
\caption{Handling a reflective boundary by flux redirection}
\end{figure}\label{f4}
right half of the tooth need be simulated since the assumption is that
the left half is its mirror image and the external boundary can be
handled as reflecting.  However, we still have to get a value for
${\rm I}_{l,0}$ since for internal boundaries it would depend in part on the
unknown  ${\rm O}_{l,0}$.  However, the reflecting condition means
that we can use ${\rm O}_{r,0}$ for the value of ${\rm O}_{l,0}$.
this is simply handled by reflecting $(1 - \alpha)$ of the output from
the right boundary of tooth 0 back in, and sending the remaining
$\alpha$ onto tooth 1 as part of its right-going input to its left
boundary, as shown in the lower part of Figure 4.
Note that this external boundary scheme also preserves the number of
particles and other
conserved values carried by the particles at this boundary.

\subsection{A Trivial Example}

Consider the linear equation
\eqn
\frac{\partial u}{\partial t} + \frac{\partial u}{\partial x} = 0
\enn{de1}
where $u(x,t)$ is a smooth approximation to the local ``density'' of a
collection of particles.  While we shouldn't be using computer time
on such a simple problem, we can simulate it by moving each particle a
distance $\Delta$ to the right in each time step of length $\Delta$.
Now suppose that we are doing this using teeth that occupy a fraction
$\alpha$ of the spatial domain.  For convenience, consider a time step
of size $\alpha D$.  (Other time steps will  give the same result, but
would require a more complex discussion.)  Hence, in one time step,
all of the particles in each tooth move from their current positions
to positions $\alpha D$ to the right - outside of the tooth they came
from.  Hence, $\alpha$ of them should move to the neighboring tooth to
the right - by being moved a further distance $(1 - \alpha)D$ - and
$(1 - \alpha)$ of them should be moved back to the originating tooth
- by being moved left an amount $\alpha D$.  (If the time step were
longer than $\alpha D$ some of the particles so moved would still be
outside of the destination tooth and would have to be subject to the
same rule again.  In effect, a larger time step would be handled as a
number of time steps of length $\alpha D$ followed by one shorter time
step.)

Let us look at the equation for the particle density, where the
density is simply approximated as a constant over the tooth and
determined by the number of particles in the tooth (modulo some
scaling which is not important in this example.)  We will denote the
density in tooth $i$ and time step $n$ as $u_{i,n}$.  We have (within the
rounding resulting from the fact that we are dealing with integral
numbers of particles)
\eq
u_{i,n+1} =  (1 - \alpha)u_{i,n} + \alpha u_{i-1,n}
\en
or
\eq
u_{i,n+1} - u_{i,n} = -\alpha (u_{i,n} - u_{i,n-1})
\en
which is precisely an upwind finite difference approximation to eq. \r{de1}
with $\Delta t/\Delta x = \alpha$ which, since $\Delta x = D$, is
consistent with $\Delta t = \alpha D$. 

\section{Higher-order Boundary Conditions and ``Anti Particles''}

The method discussed above uses linear interpolation.  This is
adequate in some cases, but not others.  Consider the heat equation
\eqn
\frac{\partial u}{\partial t} = \nu \frac{\partial^2 u}{\partial x^2}
\enn{de2}
This is the asymptotic solution of a simulation of a number of particles
subject to a suitable random walk as the number of particles goes to
infinity.  We could (if we wanted to waste computer cycles) simulate
it by taking a collection of particles in space and, at each of a
sequence of time steps, taking half of the particles (picked at random
or by some means that doesn't bias the result such as every other one
in spatial ordering) and moving them left a distance
$r$ and moving the other half right the same distance.  The distance they
should be moved is proportional to the square root of $\Delta t$ and
$\nu$, namely
\eqn
r = \sqrt{2\nu \Delta t}
\enn{r1}
so that $r$ is the standard deviation of the distribution.
(With this ``binary'' walk the asymptotic convergence to the solution
of the heat
equation is also in terms of the time step going to zero.  If each
particle were subject to an independent Gaussian movement with mean
zero and standard deviation $r$, the result would be independent of
the time step, but we will use the binary walk for ease of
discussion.)

What happens if we use the scheme from the previous section and apply
the simple analysis above to it?  Again, let us choose a time step
such that $r = \alpha D$.  In each time step, half of the particles
will exit a tooth to the left and half to the right.  If we use the
flux redirection scheme based on linear interpolation we will get the
approximate equation for $u_{i,n}$:
\eq
u_{i,n+1} = (1 - \alpha)u_{i,n} + \alpha (u_{i-1,n} + u_{i+1,n})/2
\en
which is a finite difference approximation to eq. \r{de2} if $\alpha =
2\nu\Delta t/\delta x^2$, but this requires that
\eqn
r = \alpha D = 2\nu \Delta t/D
\enn{r2}
This gives a different value of $r$ that that given in eq. \r{r1}.
The reason is that linear interpolation is not adequate for the
boundary condition in this case.  Diffusion is dependent on the {\em
derivative of the flux}, not on its actual level, so we need at least
a second-order interpolant.

Referring back to Figure 3 we could do quadratic interpolation on
the three values ${\rm O}_{r,i}$, $i = 0, 1, 2$ to estimate either
${\rm I}_{r,1}$  or ${\rm I}_{r,2}$.  Using the former, we have
\eqn
{\rm I}_{r,1} = \frac{\alpha(1+\alpha)}{2} {\rm O}_{r,0}+(1 -
\alpha^2) {\rm O}_{r,1}-\frac{\alpha(1 - \alpha)}{2} {\rm O}_{r,2}
\enn{i2}

Note that the coefficient of the last term is negative.  Any
higher-order interpolant has to have some negative coefficients (which
accounts for their poor properties near sharp fronts).  We
note that with the quadratic interpolation in eq. \r{i2} the following
fractions of the output ${\rm O}_{r,1}$ should be sent to the inputs:
\begin{itemize}
\item $(1 - \alpha^2)$ to ${\rm I}_{r,1}$
\item $\alpha(1+\alpha)/2$ to ${\rm I}_{r,2}$
\item $-\alpha(1-\alpha)/2$ to ${\rm I}_{r,0}$
\end{itemize}
This raises the issue of how to move a negative number of particles
from the output of one tooth to the input of a neighboring
tooth.  We propose sending ``anti particles,'' namely, particles whose
state is the negative of the state of a regular particle, and which
will annihilate a regular particle later.  (In the heat equation model,
the only state of a particle is its existence - which could be viewed
as its unit mass.  If a particle also had other state variables such
as momentum, the anti particle would have the same velocity, but
because of its negative ``mass'' its momentum would be the reverse.
In the simulations we have done, an anti particle is annihilated by the
nearest particle after the completion of the time step.  If it also
carried momentum, it
would be necessary to move any excess or deficit momentum after the
mutual self destruction of two particles to another neighboring
particle.)
With this model, we can use direction of right-going output flux in
second order interpolation as shown below:
\begin{itemize}
\item Send $(1 - \alpha(1+\alpha)/2)$ of the particles to the input of
the same tooth 
\item Send $\alpha^2$ of the particles to the input of the right neighbor
\item Send the remaining $\alpha(1-\alpha)/2$ of the particles to the right
neighbor, send copies of them to the same tooth, and send
anti-particle copies of them to the input of the left neighbor.
\end{itemize}
Note that the sum of the number of particles (where anti-particles
are negative in the count) is unchanged so there is preservation of
``mass'' and other conserved values.

We now return to the heat equation example.  With $r$ equal to $\alpha
D$ as before, we now find that we get the finite difference
equation
\eq
u_{i,n+1} = (1 - \alpha^2)u_{i,n} + \alpha^2 (u_{i-1,n} + u_{i+1,n})/2
\en
which is the finite difference approximation to the heat equation with
$\alpha^2 = 2\nu\frac{\Delta t}{\Delta x^2}$.  Noting that $r = \alpha
D$ and $\Delta x = D$ we see that
\eq
r = \sqrt{2\nu\Delta t}
\en
agreeing with eq. \r{r1}.

\section{Comments}

The key ideas in this note are the redirection of the outward fluxes
based on the spacing of the teeth, and the use of ``anti particles''
to handle negative interpolation coefficients.  The ideas go over to
higher dimensions in which a particle may exit a tooth from one
or several boundaries.  All that is necessary is to do the redirection
in each space dimension separately, so that the particle could move to
any one of the $3^n -1$ nearest neighbors in $n$ dimensions (or more
distant neighbors if even higher order interpolation is needed). 
While we have conserved the number of particles with the schemes
discussed, a wider range of schemes can be considered if this feature
is dropped.  Dropping conservation may be necessary to handle a
non-uniform mesh
(although are ways to maintain conservation by using more inputs in an
interpolation than necessary, but this
issue has not been fully explored).


\newpage   

\begin{thebibliography}{99}

\bibitem{gaptooth} Kevrekidis, I. G., Coarse Bifurcation Studies of
Alternative Microscopic/Hybrid Simulators, Plenary Lecture, CAST Division
of the AIChE, AIChE Annual Meeting, Los Angeles, 2000, slides at
http://arnold.princeton.edu/~yannis/

\bibitem{eqfr}Equation-Free Multiscale Computation: enabling
microscopic simulators to perform system-level tasks,  NEC research
Institute Report 2002-010N, Aug, 2002, Submitted to Communications in the
Mathematical Sciences (with  I. G. Kevrekidis, J. M. Hyman,
P. G. Kevrekidis, O. Runborg, and C.Theodoropoulos) at
http://www.neci.nj.nec.com/homepages/cwg/

\end{thebibliography}
\end{document}